\documentclass[conference]{IEEEtran}
\IEEEoverridecommandlockouts

\usepackage{cite}
\usepackage{amsmath,amssymb,amsfonts}
\usepackage{algorithmic}
\usepackage{graphicx}
\usepackage{textcomp}
\usepackage{xcolor}
\usepackage{paralist}
\usepackage{subcaption}

\def\BibTeX{{\rm B\kern-.05em{\sc i\kern-.025em b}\kern-.08em
    T\kern-.1667em\lower.7ex\hbox{E}\kern-.125emX}}
\begin{document}

\title{\Large\textbf{ Hardware Versus Software Fault Injection of Modern Undervolted SRAMs}}

\author{\IEEEauthorblockN{Muhammet Abdullah Soyturk\IEEEauthorrefmark{1},
Konstantinos Parasyris\IEEEauthorrefmark{2}, 
Behzad Salami\IEEEauthorrefmark{2},
Osman Unsal\IEEEauthorrefmark{2},\\
Gulay Yalcin\IEEEauthorrefmark{1} and
Leonardo Bautista Gomez\IEEEauthorrefmark{2}}
\\
\IEEEauthorblockA{
\IEEEauthorrefmark{1}\textit{\{muhammetabdullah.soyturk, gulay.yalcin\}@agu.edu.tr}\\
\textit{Abdullah Gul Univ, Kayseri, Turkey}\\
\IEEEauthorrefmark{2}\textit{\{konstantinos.parasyris, behzad.salami, osman.unsal, leonardo.bautista\}@bsc.es}\\
\textit{Barcelona Supercomputing Center, Barcelona, Spain}
}}

\maketitle

\begin{abstract}
To improve power efficiency, researchers are experimenting with dynamically adjusting the supply voltage of systems below the nominal operating points. However, production systems are typically not allowed to function on voltage settings that is below the reliable limit. Consequently, existing software fault tolerance studies are based on fault models, which inject faults on random fault locations using fault injection techniques. In this work we study whether random fault injection is accurate to simulate the behavior of undervolted SRAMs.

Our study extends the Gem5 simulator to support fault injection on the caches of the simulated system. The fault injection framework uses fault maps, which describe the faulty bits of SRAMs, as inputs. To compare random fault injection and hardware guided fault injection, we use two types of fault maps. The first type of maps are created through undervolting real SRAMs and observing the location of the erroneous bits, whereas the second type of maps are created by corrupting random bits of the SRAMs. During our study we corrupt the L1-Dcache of the simulated system and we monitor the behavior of the two types of fault maps on the resiliency of six benchmarks. The difference among the resiliency of a benchmark when tested with the different fault maps can be up to 24\%. 
\end{abstract}

\begin{IEEEkeywords}
Fault Injection, Fault Models, Voltage Underscaling, SRAMs
\end{IEEEkeywords}

\section{Introduction}

As predicted by Moore\textquotesingle s Law, the scalability of semiconductor manufacturing process has been the driving force behind the increase in the capabilities of computer systems. However, scaling in lower nanometer geometries has led to variability of transistor characteristics, resulting into increased failure rates in modern CPUs. Conventional techniques for providing reliable execution include extra provisioning in logic and memory circuits in the form of increased voltage margins and reduced operating frequencies (so-called guardbands), as well as special error correction circuitry. But all these techniques consume more power, thus they are not very attractive in light of the ambitious goal to reach exascale performance with constrained power budgets. More specifically, guardbanding may increase power dissipation in the order of 35\% ~\cite{1610623}. 

A promising way to increase the energy efficiency of modern systems, is to remove these guardbands by reducing the supply voltage of the system while keeping the operating frequency constant. By doing so, the energy/power consumption of the system is decreased on average by  $20\%$ ~\cite{Papadimitriou:2017:HVM:3123939.3124537,Parasyris:2017:SPE:3086564.3058980}. One of the key aspects of this technique is that the performance remains the same, since frequency remains the same, all the energy saving results from the  power reduction of the system. 

On the one hand, there are many studies which perform supply voltage underscaling on real systems and monitor the system. These studies either focus on very specific architectures and technologies, for example SRAMs \cite{8410585,Abella:2009:LVF:1669112.1669128,8671543} or they perform voltage underscaling on the entire system and monitor the system as a black box \cite{Papadimitriou:2017:HVM:3123939.3124537,8416495}. In the first case, the studies provide very accurate descriptions of the fault locations and the timings of the faults, however the specific observations are limited to the specific system and technology. On the second case, there is almost no information on the actual fault rate as errors are observed in the output of the software, thus masked faults are not captured at all. 

On the other hand, there are studies which implement software based fault tolerance techniques which try to yield the energy benefits of voltage underscaling and will handle potential errors at the software level \cite{Parasyris:2017:SPE:3086564.3058980,Achour:2015:ACO:2814270.2814314}. To evaluate these techniques, they employ fault injection mechanisms which are guided by fault models and predicts the probability of a fault for a specific voltage setting. 
Fault models typically, uniformly distribute faults to all faulty locations. The uniform distribution of errors is not accurate. For example, when underscaling the supply voltage of SRAMs, spatially related errors are common \cite{8671543,8574581}.

In this paper, we compare the accuracy of random based fault injection with the accuracy of hardware guided fault injection when undervolting the SRAM of a system. We implemented a fault injector framework which uses fault maps as inputs in Gem5 \cite{Binkert:2011:GS:2024716.2024718}. The faults are injected during the execution of the
simulation. In the end we observe the manifestation of faults to the
applications output. The fault maps are created using 
\begin{inparaenum}
\item A fault model in which the number of injected faults is guided by the number of
faults appearing on different real SRAMs, but  the faulty location is randomly selected \cite{8060425}. 
\item Actual fault maps  publicly available from ~\cite{github}. These fault maps describe the number of errors as well as their specific bit location of actual undervolted SRAMs.
\end{inparaenum}

The main contributions of this work are the following:
\begin{itemize}
    \item We extend the Gem5 simulator to support fault injection on the different caches of the system. The fault injection framework uses as input a fault map, which describes the location of the faults. These faults are injected during the execution of the simulation
    \item We evaluate two fault types of fault maps, the first fault maps are created with random fault injection fault model, whereas the second are fault maps created by undervolting real SRAMs and capturing the location of the errors. 
\end{itemize}

The rest of the paper is structured as follows. In Section \ref{sec:motivation}
we motivate our study. Section \ref{sec:methodology} describes the methodology
we used to perform our analysis and study. In Section \ref{sec:evaluation} we
present our evaluation. Section \ref{sec:related} outlines the related work and
Section \ref{sec:conclusions} concludes the paper. 

\section{Motivation}
\label{sec:motivation}
In this paper, we study whether random fault injection guided only by a fault rate is sufficient to capture the realistic fault manifestation of undervolted errors. Evaluating fault models is of great importance as the coverage of any software error detection technique actually depends on the fault model. On the one hand, spending resources to detect errors which will never appear is not efficient. On the other hand understanding the actual manifestation of undervolted errors on the software/architecture level can result to more accurate error detection mechanisms.

Fault models guide the procedure of fault injection, which is the typical method to evaluate the robustness of a system. Depending on the approach different fault injections techniques can be utilized. For example, when trying to detect errors during the execution of the application, researchers typically use random  based fault injection techniques. This random fault injection is usually guided by a fault model which determines the number of errors to inject in each experiment ~\cite{8060425,Bpredictor,Parasyris:2017:SPE:3086564.3058980} depending on the supply voltage. These models are agnostic to the actual hardware fault location. Since their techniques are agnostic to the underlying architecture and thus they should detect a wide range of errors without focusing on how these faults are propagated to the software level.  

However, a specific SRAM block will behave deterministically when applying voltage underscaling. In other words, the errors will appear on deterministic fault locations for the specific SRAM and the specific supply voltage ~\cite{8574581}. Although, the same SRAM structure on different chips will present different fault locations, all of the studied undervolted SRAMs present a spatial error locality ~\cite{8671543}. In other words, bits that are in close proximity from a faulty bit have higher probability to be also faulty. Moreover, when undervolting an SRAM structure there exist a \textit{fault inclusion} property ~\cite{8671543}. Namely, when a bit is faulty for an undervolted supply voltage setting $V_{u}$ it will be faulty for any supply voltage $V_{x} < V_{u}$. This behavior is again not captured by random based fault injection methods . 

\begin{figure}[!t]
    \centering
    \begin{subfigure}[t]{0.22\textwidth}
        \centering
        \frame{\includegraphics[width=\textwidth]{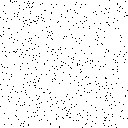}}
        \caption{Random fault distribution of an 16Kbits SRAMs.}

    \end{subfigure}
    \hfill
    \begin{subfigure}[t]{0.22\textwidth}
        \centering
        \frame{\includegraphics[width=\textwidth]{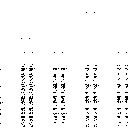}}
        \caption{Fault distribution from a real undervolted 16Kbits SRAMs}
        
    \end{subfigure}
    \caption{Figures represent errors using a random fault model, and a realistic fault map. Each pixel of the image corresponds to a specific bit of the SRAM. These fault maps present 600 faulty bits (represented in black color) out of the total 16Kbits.}
    \label{fig:errorDistribution}
\end{figure}

In Figure \ref{fig:errorDistribution} we present two different fault maps, the first one in Figure \ref{fig:randomErrors} is created by using a random fault model, the second one, in Figure \ref{fig:realErrors}, is a fault map of an actual publicly available SRAM at ~\cite{github}. Noticeably, although both fault maps present the same number of errors, the faulty locations are different. Namely in the random fault map, each bit has the same probability to manifest an error, therefore, all errors are distributed through the entire structure. On the other hand, on actual hardware, a few errors exist in the higher part of the figure, whereas most of them are clustered in the bottom of the figure. Interestingly, most of the errors are gathered in a column wise manner, in other words when an error appears in a specific column, there are many errors on the same column on lower lines ~\cite{8671543}. 

\section{Methodology}
\label{sec:methodology}

Our objective is to identify whether the hardware fault location matters from the perspective
of the software when undervolting the SRAMs 
of the system. The methodology 
consists of 3 steps. The first step is the creation of fault maps. A fault map is 
a description of which bit or bits are faulty for a specific SRAM. Each generated 
fault map is given as input to a fault injection framework. The fault injection 
framework, is the second step of our methodology. It is based on Gem5 and simulates a faulty  system. During the simulation, it corrupts the respective bits in the cache as 
described in the fault map. The procedure performs a single simulation for each 
fault map. The fault injection framework can inject errors to any cache of the 
system. Since simulation based fault injection is a timing consuming procedure, in this work we opt to study only the effect of faults corrupting the L1-DCache and not the remaining cache levels.  The third step corresponds to the classification of the output. After each simulation 
we compare the output of the application with a \textit{golden} output, an output that is 
created by the application when executed with no errors. Depending this comparison the 
experiments are categorized into different categories.
In the following sections we provide more detailed information for each of this
steps. 

\subsection{Fault Map Generation}

\begin{figure}
\includegraphics[width=0.5\textwidth]{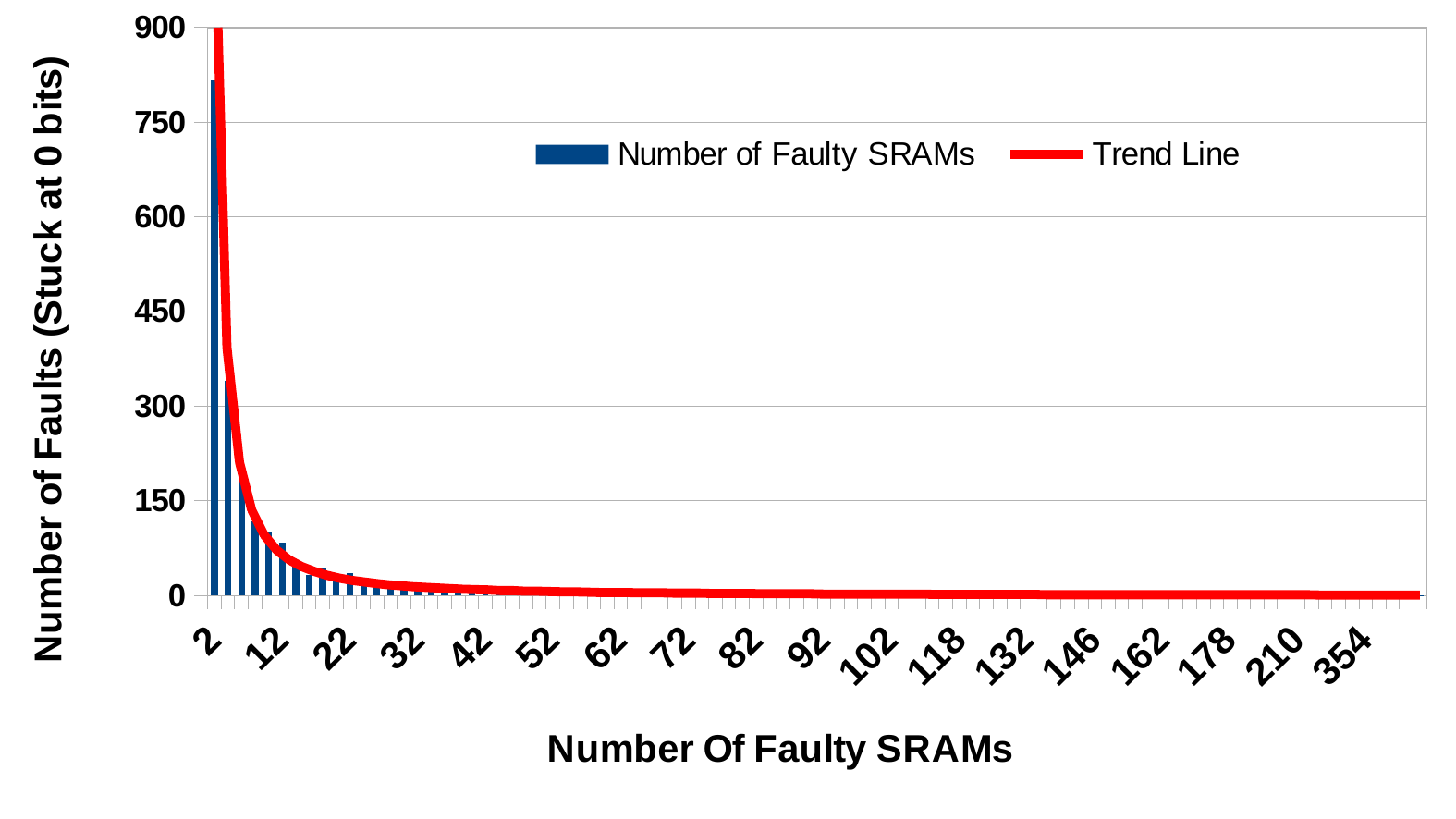}
\caption{Number of faulty SRAMs (X-axis) that manifest different number of stuck-at 0 faults (Y-axis)} 
\label{fig:distribution}
\end{figure}

The hardware guided fault injection fault maps are provided in ~\cite{github}. There exist different fault maps for a SRAM structure of $28nm$ for $7$ different undervolted supply voltage settings, $0.54V - 0.60V$ with a step of $0.01V$. For each of these settings there are $2060$ individual fault maps, each one of them represents a single SRAM structure. Each SRAM is of size $16Kbits$. There exist in total $2060*7=14420$ different fault maps. From these fault maps only the $2174$ maps manifest errors during the undervolting. The errors are stuck-at-0 ones, as previous studies show that the majority of undervolting SRAMs errors are stuck-at-0 ~\cite{8374828,8671543}. The number of errors in each fault map differs depending on the specific map. In Figure \ref{fig:distribution} we present from the faulty SRAM structures how many of them (Y-axis) exist with a specific number (X-axis) of errors. As depicted in Figure \ref{fig:distribution} most of the SRAMs structures present a small number of faults, namely $37\%, 15\%, 9\%$ and  $5\%$ of the faulty SRAMs present $2, 4, 6, 8$ of fault bits respectively.

The random fault injection fault maps are generated with the following methodology. For each of the faulty hardware faults maps we create an equivalent fault map in terms of number of faults. For example if the hardware fault map contains $4$ faults we create a random fault map with also $4$ faults. However, the location of the faults are distributed randomly in the bits of the SRAM. By doing so we can compare only the effect of the faulty location to the applications resiliency while preserving the error rate exactly the same among the two methods. From now on the hardware guided fault injection will be referred ad \textit{HW FI} whereas the random one as \textit{RND FI}.

\begin{figure}[!tb]
    \centering
    \begin{subfigure}[t]{0.45\textwidth}
        \centering
        \includegraphics[width=\textwidth]{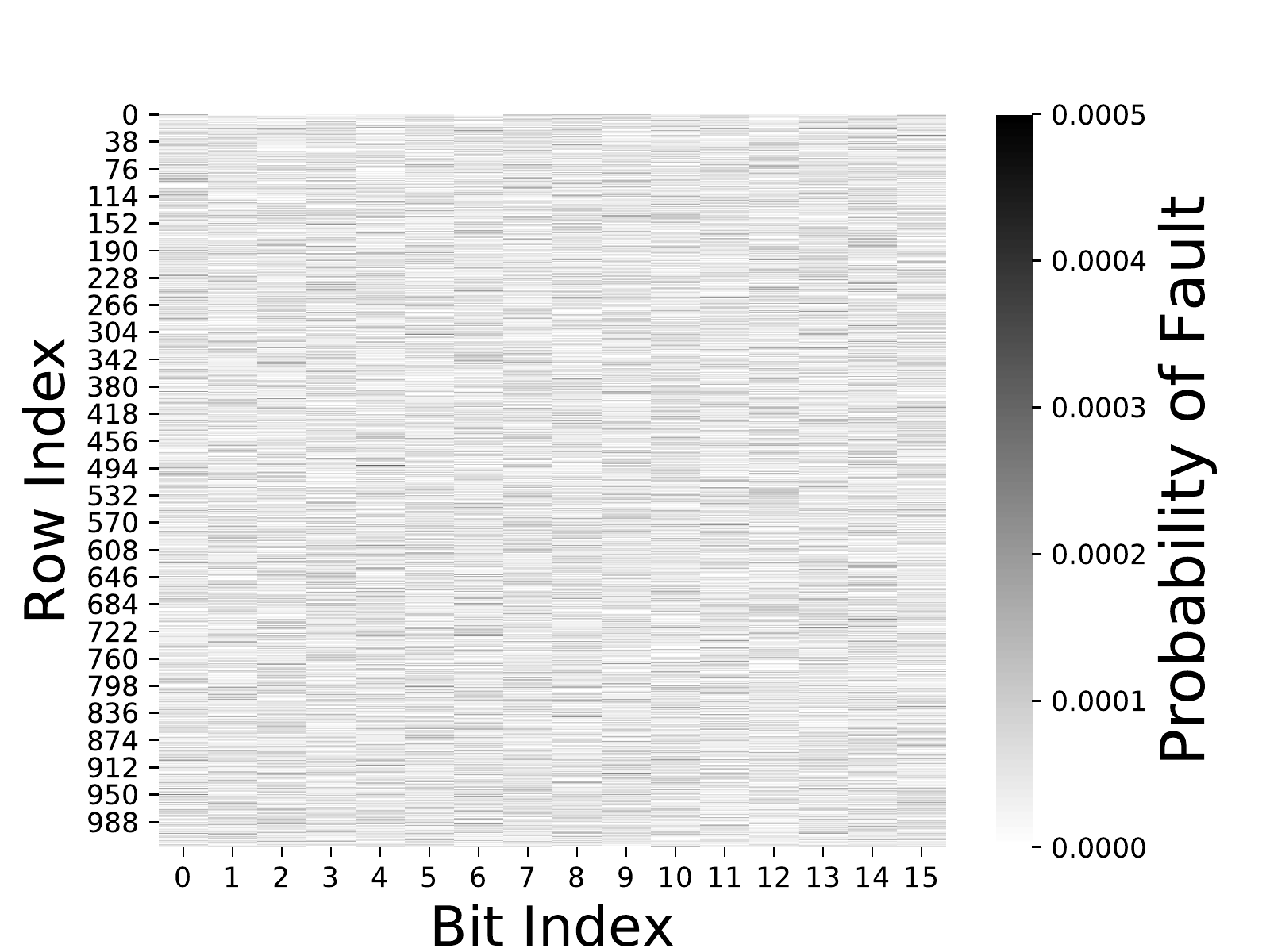}
        \caption{Probability of error for every bit using \textit{RND FI}.}
        \label{fig:randomErrors}
    \end{subfigure}
    \hfill
    \begin{subfigure}[t]{0.45\textwidth}
        \centering
        \includegraphics[width=\textwidth]{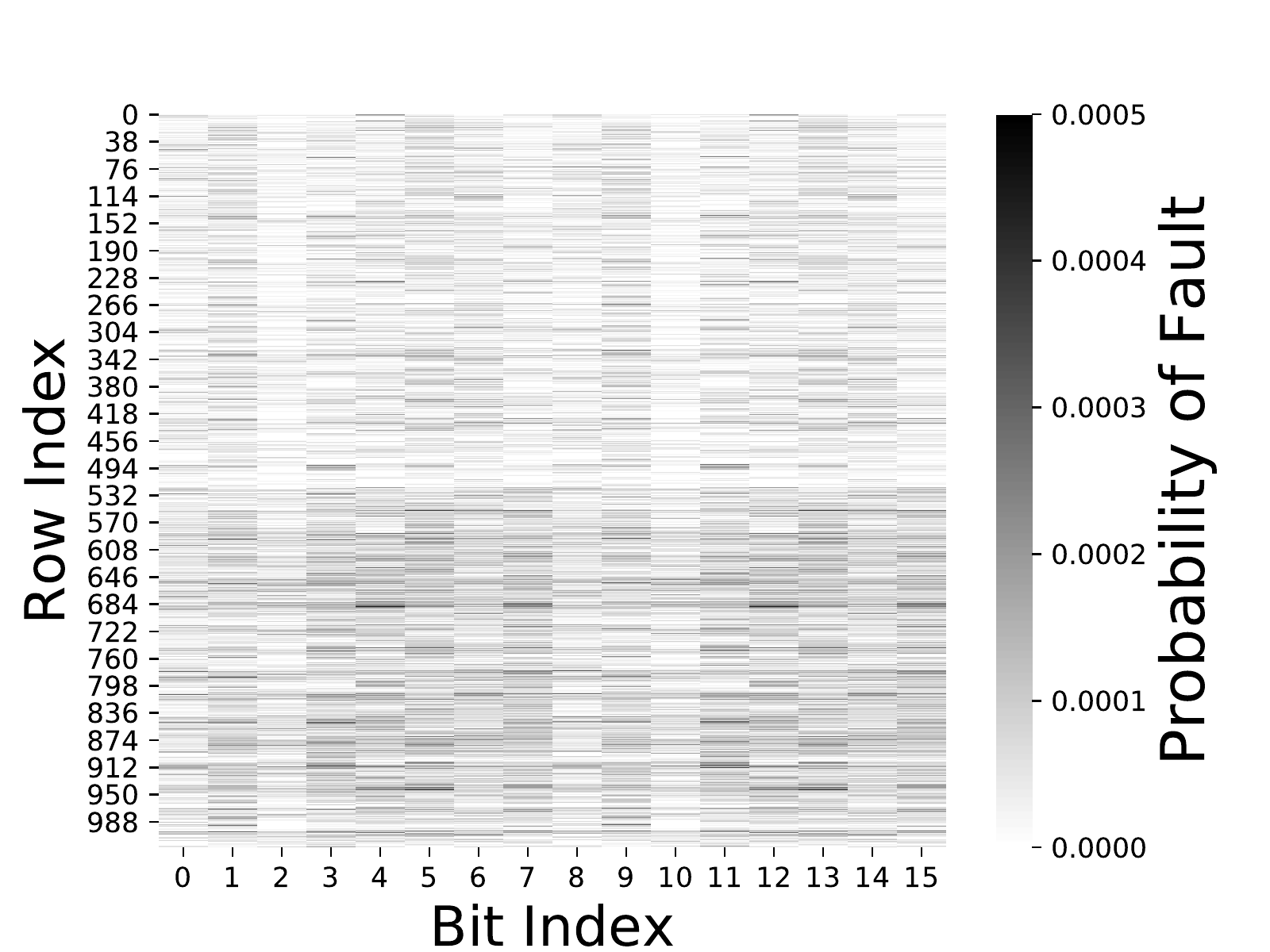}
        \caption{Probability of error for every bit using \textit{HW FI}.}
        \label{fig:realErrors}
    \end{subfigure}
    \caption{Figures representing the probability of an error to be faulty.}
    \label{fig:faultMaps}
\end{figure}

In Figure \ref{fig:faultMaps} we depict the probability of a fault to occur on a specific bit using the two fault map generation techniques. As it is obvious in the \textit{RND FI} all bits have the same probability whereas in the \textit{HW FI} bit on the higher addresses are more probable to manifest errors. 

\subsection{Fault Injection Framework}

The fault injector framework uses a fault map, which was provided with one of the methods described above, as an input. During the initialization of the simulator the fault map is read and the framework creates a C++ object for each of the faults. A fault is described by the location, the timing, and corruption type. The location is described in a hierarchical order, type of cache (D / I cache), the cache level (L1, L2 etc), which byte to corrupt in the  cache line and which bit in the byte. The timing behavior of the fault can be expressed as  a transient, intermittent or permanent. Transient faults are injected once on a user defined simulation tick, intermittent ones are described by the duration of the fault and the initial simulation tick in which the fault start to appear. The permanent faults constantly corrupt the faulty bit. Finally the corruption type describes whether to set the bit on a specific value, this simulates stuck-at-X faults, or to just perform a a bit flip. 

The fault injector framework intercepts all read/write requests for all the caches of the system.  For all the requests, the framework checks whether the cache locations targeted by the request to the cache include any faulty location in the fault map. If it does, the framework injects the faults to the faulty locations. The fault is guided by the fault type, stuck-at-0 faults set a specific bit to 0 etc. In the case of a transient fault after injecting the error the fault-object is removed from the internal data structures. In the case of permanent faults the fault is not removed. Therefore, in any upcoming request the value will be injected again, if necessary. This procedure allows us to observe the manifestation of errors to the applications output.

The fault injector framework can inject faults to any system simulated with the Gem5 simulator and is orthogonal to the underlying Instruction Set Architecture (ISA). In this work we focus on a \textit{X86\_64} system using system emulation mode. To project accurately the realistic SRAMs fault maps with the L1-D cache we use the same size of $16Kbits$ as the one used at the undervolting.

\subsection{Benchmark description and categorization}

To compare the random based location with hardware guided location fault models,
we used 6 different benchmarks. 
\begin{inparaenum}
\item \textit{Jacobi}:  an iterative numerical solver for determining the solutions of a diagonally dominant system of linear equations.
\item \textit{Blackscholes}:  a benchmark of the Parsec suite
\cite{Bienia:2008:PBS:1454115.1454128}. It implements a mathematical model 
for a market of derivatives, which calculates the buying and selling of assets so as to reduce the financial risk.
\item \textit{DCT}: Discrete Cosine Transform is a module of the JPEG
compression and decompression algorithm \cite{952804}.
\item \textit{MC}: applies a Monte Carlo approach to estimate the boundary of a 
sub-domain within a larger partial differential equation (PDE) domain, by performing random walks from points of the sub-domain boundary to the boundary of the initial domain
~\cite{manolisvavalis_2014}. 
\item \textit{Sobel}: a 2D filter for edge detection in images.
\item \textit{K-Means}: an iterative algorithm for grouping data points
from a multi-dimensional space into \textit{k} clusters.
\end{inparaenum}

After each fault injection experiment the output is classified as: 
\begin{inparaenum}
\item \textit{Correct}, The applications' output after injecting the faults is bitwise
exact with the \textit{golden} output. 
\item  Silent Data Corruption \textit{(SDC)}, the application terminated normally,
however, the output is not bitwise exact in comparison with the \textit{golden} one. 
\item \textit{Crash}, the application failed to terminate. 
\end{inparaenum}

The first categorization performs a classification of the faults on the resiliency of the application. However, this is not sufficient to completely characterize the effect of the faults to the quality of the application. Therefore we perform a further analysis on the quality of the output for the experiments that resulted into the \textit{SDC} category. For the visualization benchmarks \textit{DCT, Sobel} we use \textit{Peak Signal to Noise Ratio (PSNR)} as a quality metric, the  largest the value the better the quality of the output. For the numerical applications \textit{Blackscholes, MC, Jacobi} we use the \textit{Average Relative Error}. Finally, for the \textit{K-Means} we use as a quality metric the percentage of data points that were classified to the correct cluster.  The second analysis of the quality of the output is of great importance. It actually captures whether the errors had a negative impact on the quality or not. Stating that an experiment is an \textit{SDC} is not sufficient. For example, there are cases in \textit{Sobel} in which two experiments were categorized as SDCs, however, the first experiment resulted to a single wrong pixel, whereas the second just computed a black image.

\section{Evaluation and Analysis}
\label{sec:evaluation}

\begin{figure}[!t]
\includegraphics[width=0.5\textwidth]{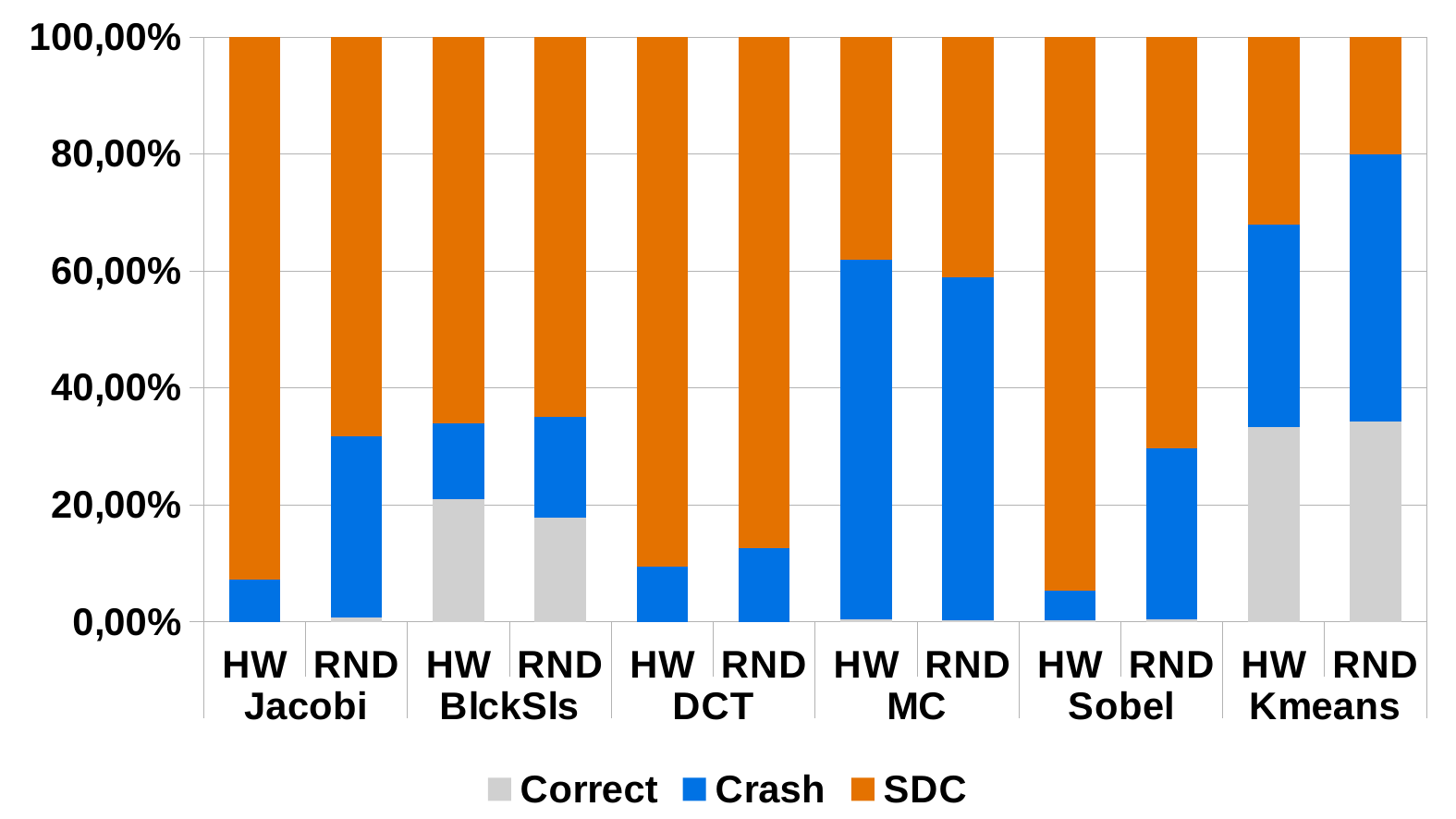}
\caption{Classification of the fault injection experiments using the two different fault injection methods.} 
\label{fig:evalClassification}
\end{figure}

In Figure \ref{fig:evalClassification} we present the high level classification of the two fault injection methodologies for all the benchmarks we studied. Interestingly 99\% of the observed crashes were due to accessing unmapped memory regions (Segmentation faults), therefore we do not provide any further classification of crashes. As depicted, three benchmarks demonstrate significant differences among the two fault injection types. Namely \textit{Jacobi, Sobel} and \textit{Kmeans} present a $ 23\%, 24\%$ and $11\%$ higher crashing rates respectively in the case of \textit{RND FI}. In the case of \textit{HW FI} the higher addresses of the cache are more often corrupted, whereas in \textit{RND FI} all addresses have equal probability to be corrupted. These three benchmarks, store pointer values to lower addresses, therefore in the case of \textit{HW FI} this values are way more infrequently corrupted, hence the decreased \textit{crash} rates. The average fault probability of the two different fault injection methods are presented in Figure \ref{fig:faultMaps}, as depicted in \textit{HW FI} the higher addresses are more probable to result into errors.   

\begin{figure}[!b]
\includegraphics[width=0.5\textwidth]{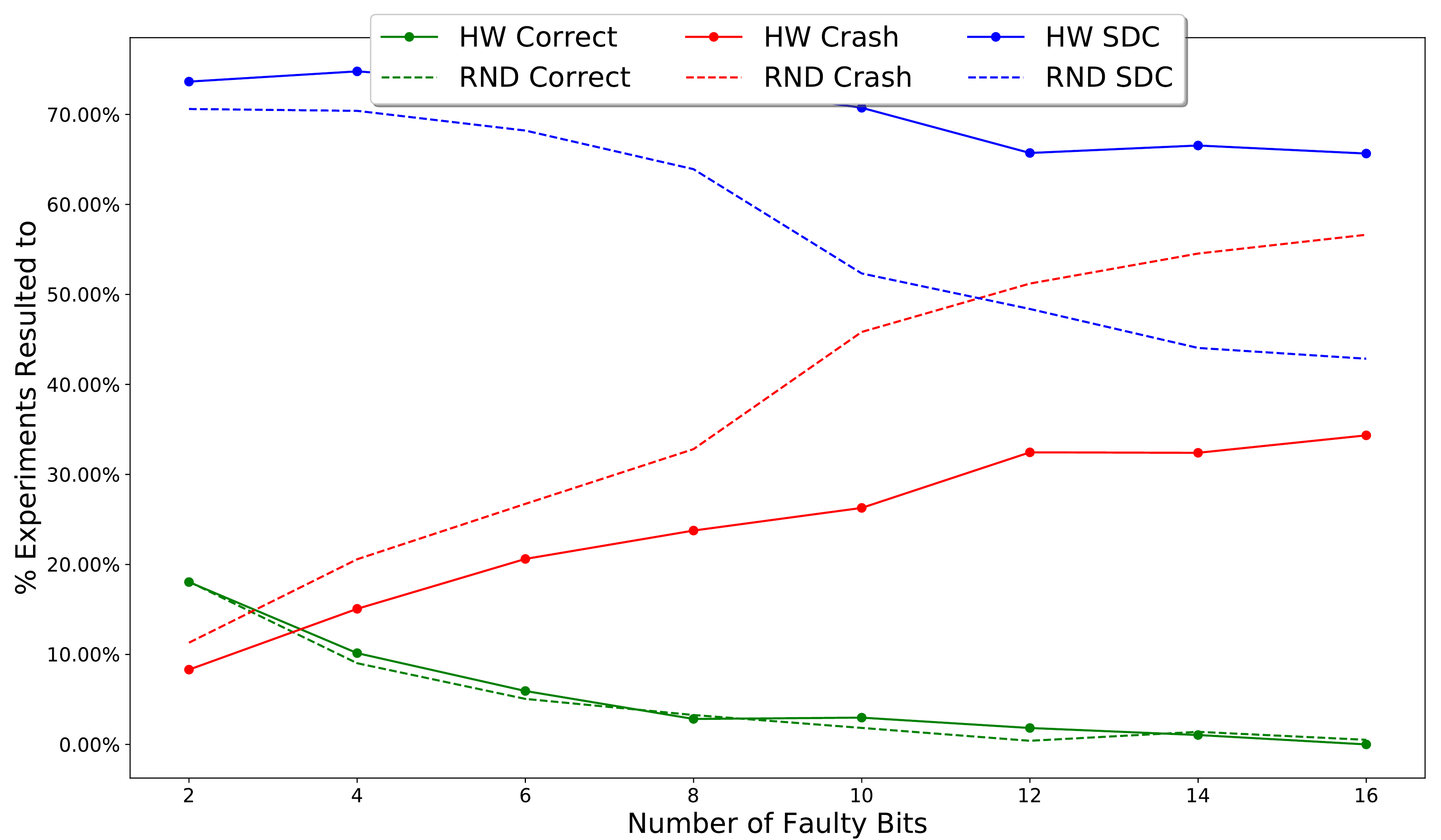}
\caption{Classification of the fault injection experiments for all the benchmarks depending on the number of faulty bits.} 
\label{fig:evalClassificationNumErrors}
\end{figure}

\begin{figure*}[!htpb]
    \centering
    \begin{subfigure}[t]{0.3\textwidth}
        \centering
        \includegraphics[width=\textwidth]{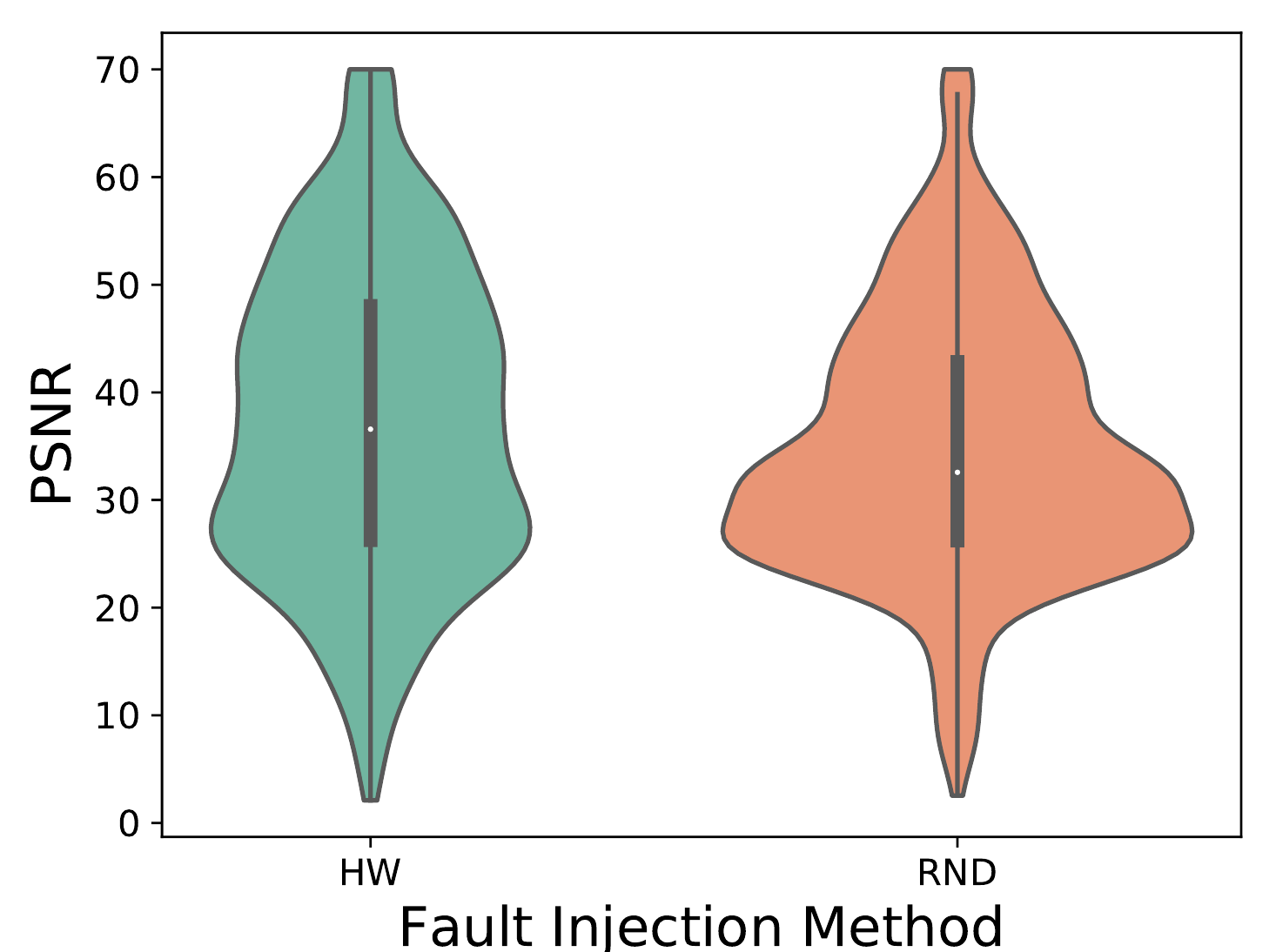}
        \caption{Sobel}
        \label{fig:sobelQual}
    \end{subfigure}
    \hfill
    \begin{subfigure}[t]{0.3\textwidth}
        \centering
        \includegraphics[width=\textwidth]{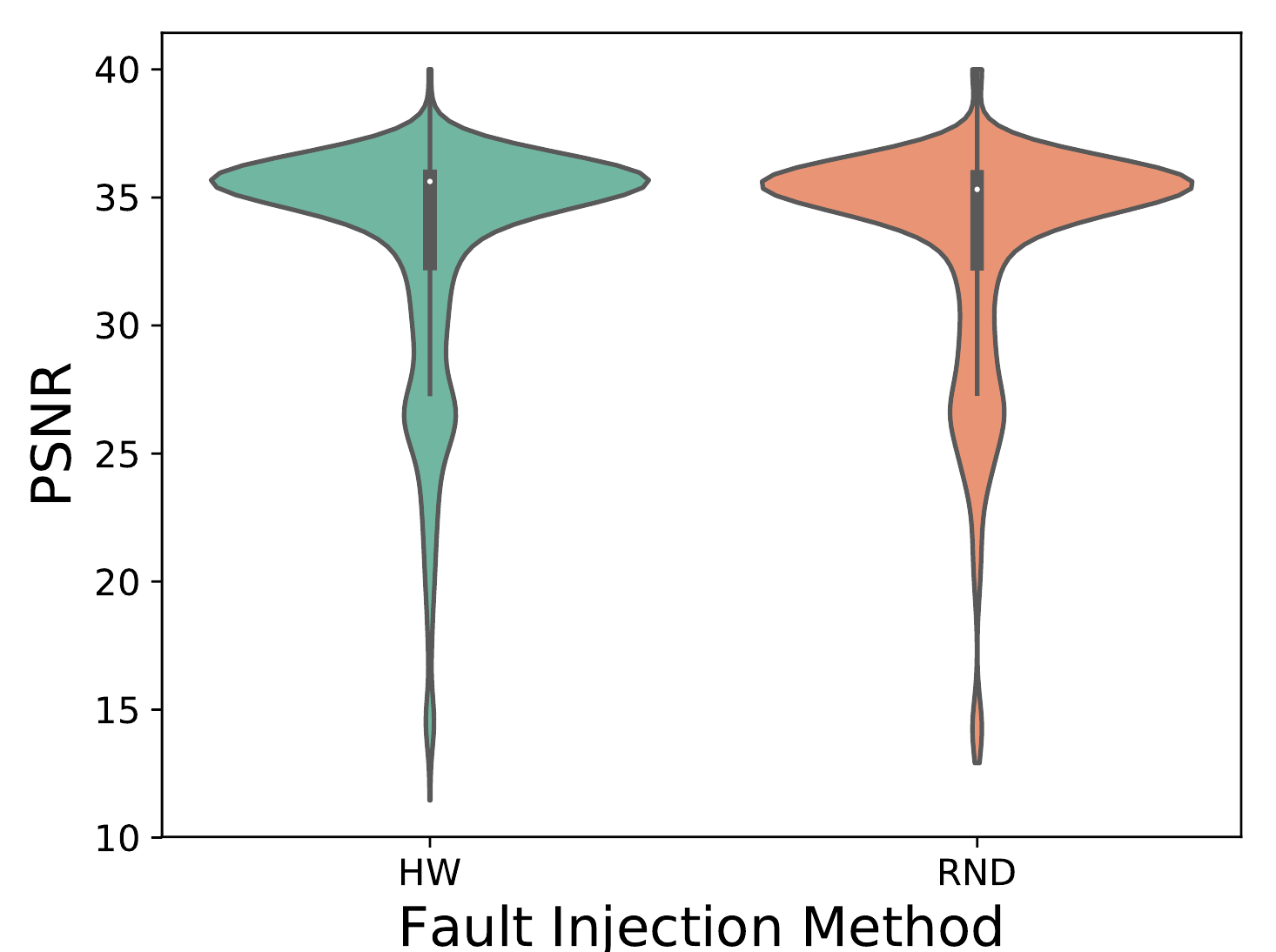}
        \caption{DCT}
        \label{fig:DCTQual}
    \end{subfigure}
    \hfill
    \begin{subfigure}[t]{0.3\textwidth}
        \centering
        \includegraphics[width=\textwidth]{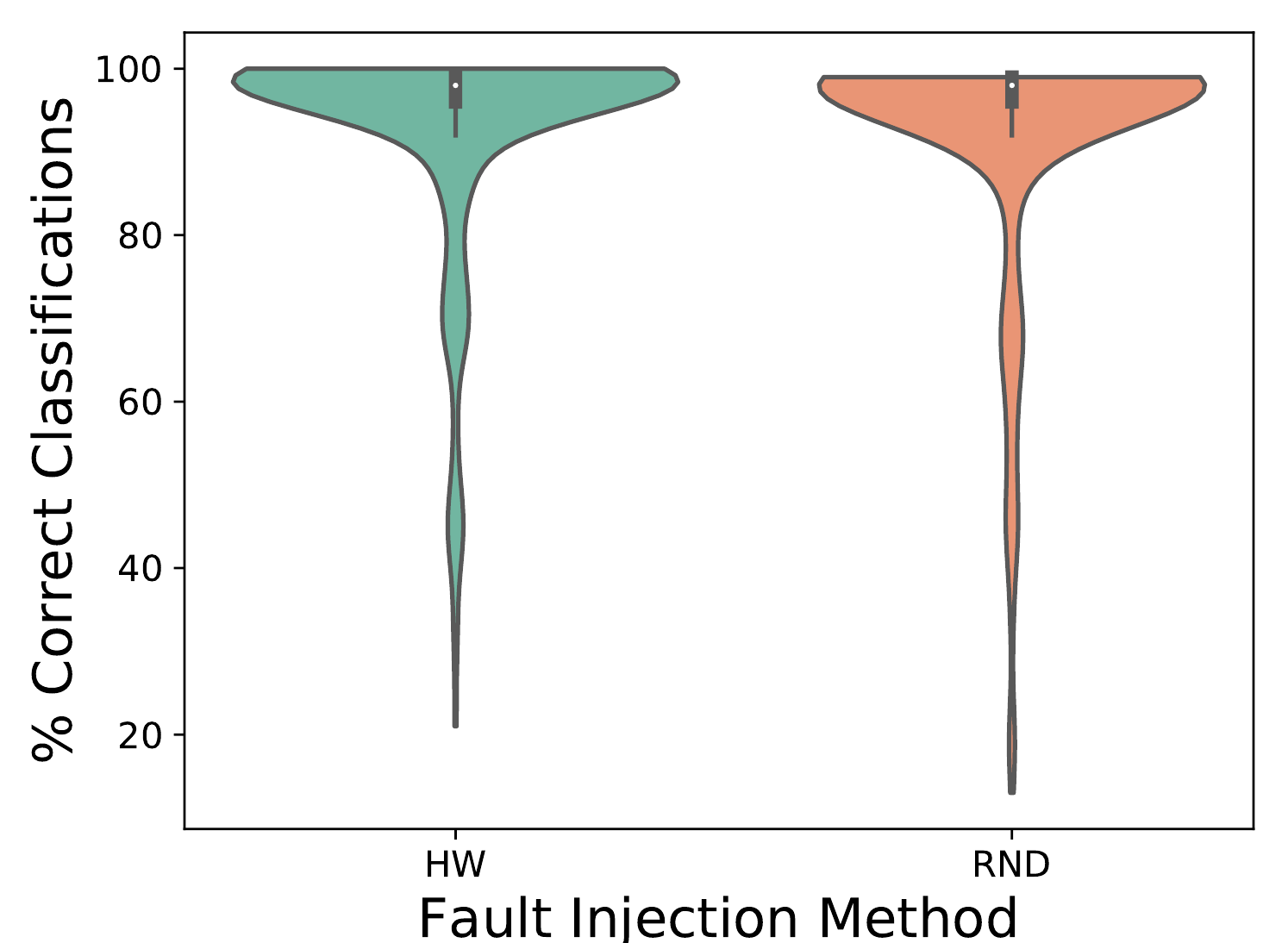}
        \caption{KMeans}
        \label{fig:Kmeans}
    \end{subfigure}
    \begin{subfigure}[t]{0.3\textwidth}
        \centering
        \includegraphics[width=\textwidth]{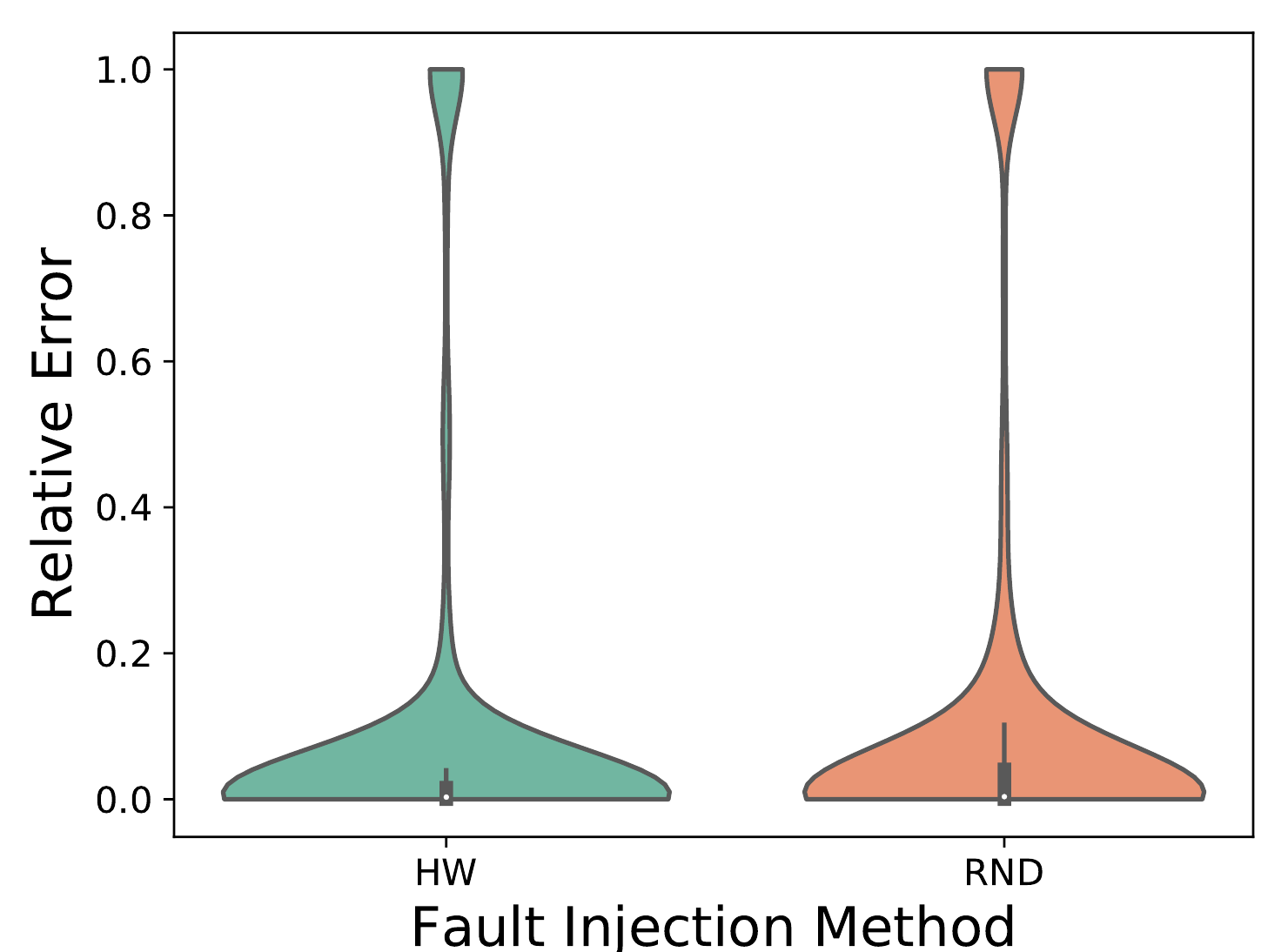}
        \caption{Blackscholes}
        \label{fig:BlackscholesQual}
    \end{subfigure}
    \hfill
    \begin{subfigure}[t]{0.3\textwidth}
        \centering
        \includegraphics[width=\textwidth]{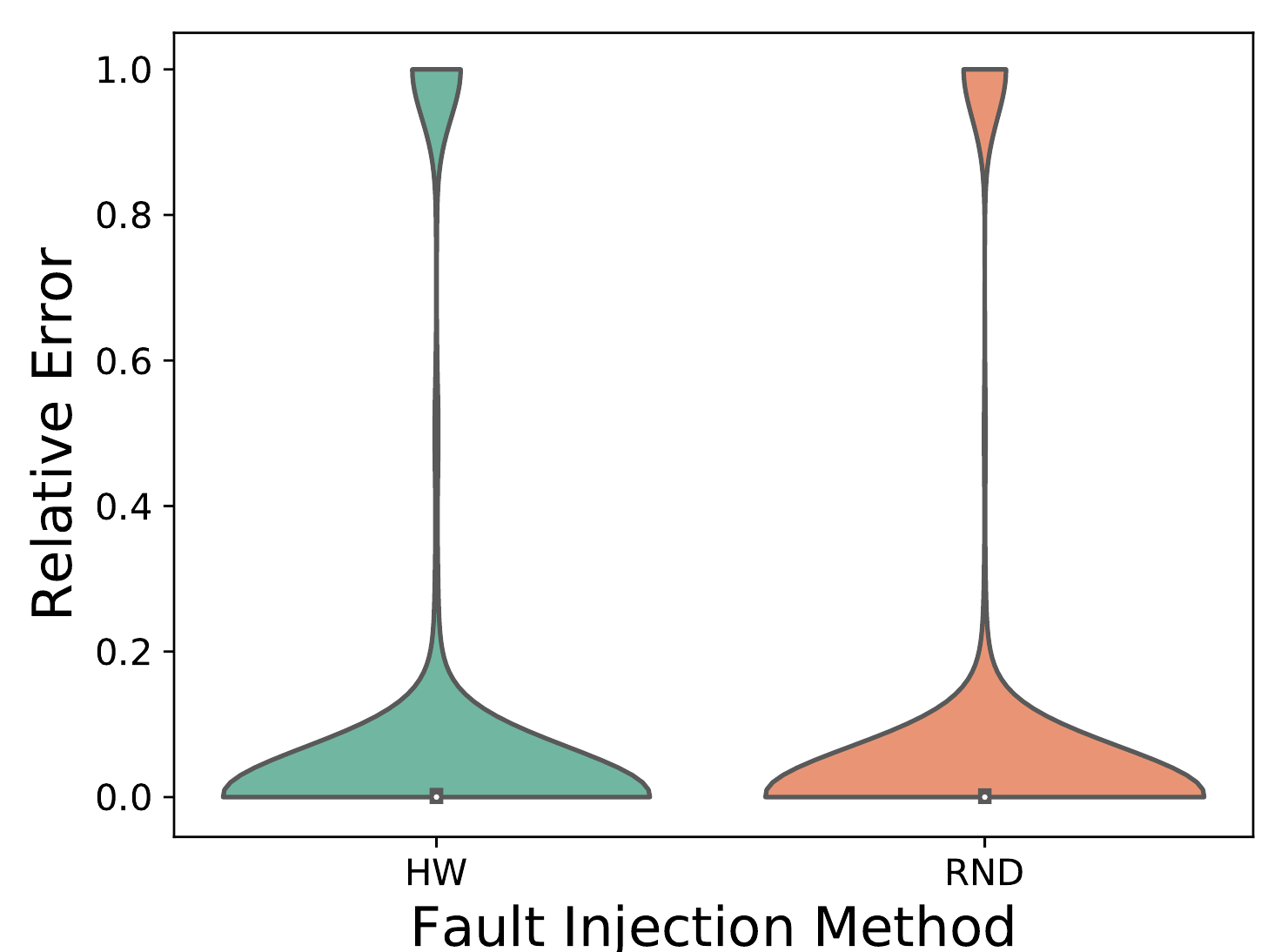}
        \caption{Jacobi}
        \label{fig:JacobiQual}
    \end{subfigure}
    \hfill
    \begin{subfigure}[t]{0.3\textwidth}
        \centering
        \includegraphics[width=\textwidth]{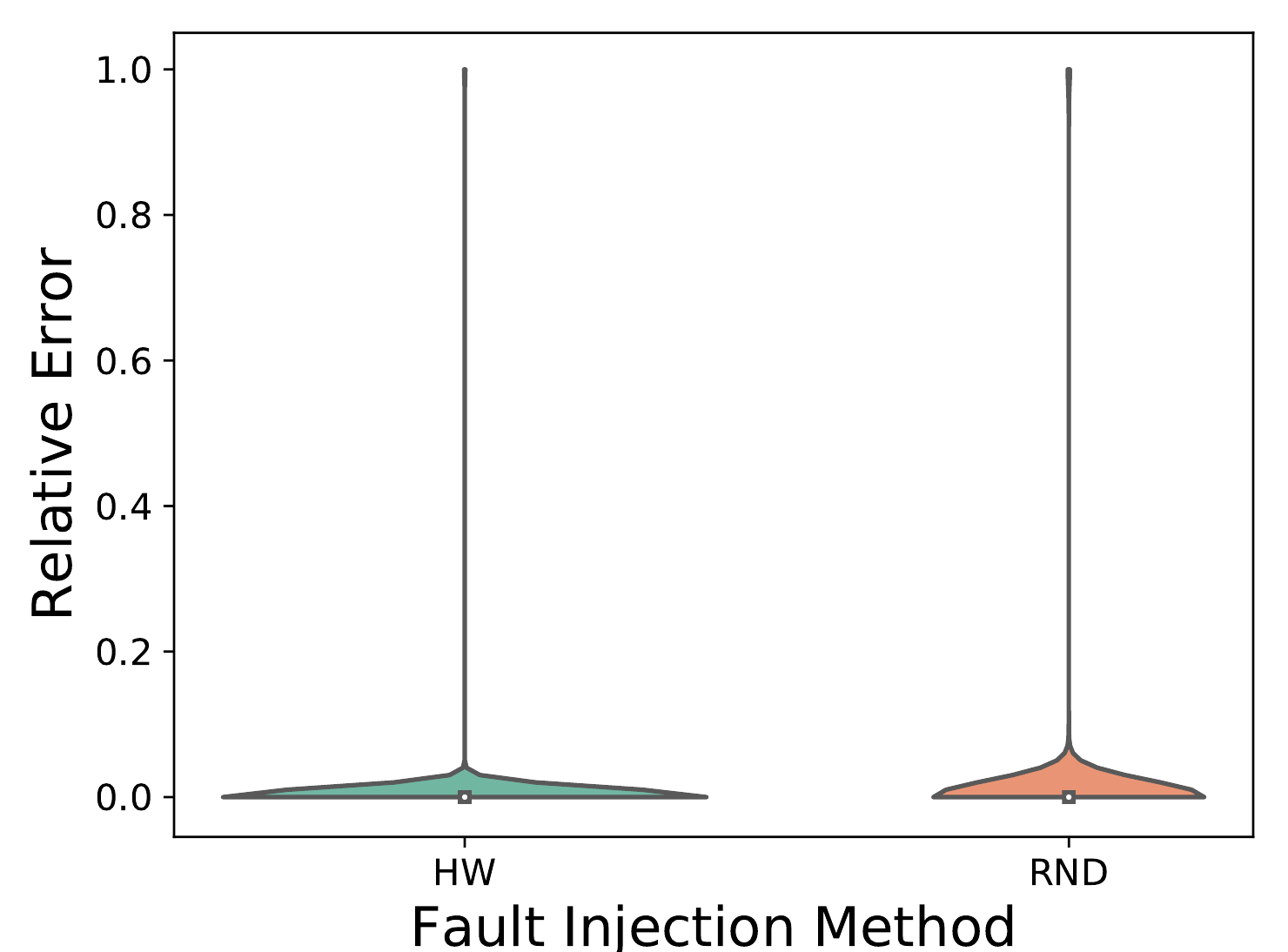}
        \caption{Monte Carlo}
        \label{fig:MC}
    \end{subfigure}
    \caption{Quality of the output for the experiments that resulted to the \textit{SDC} category.}
    \label{fig:qualities}
\end{figure*}

In Figure \ref{fig:evalClassificationNumErrors} we present the accumulated results for all the benchmarks, depending on the number of bit flips that occur per experiment. We only present experiments that corrupt up to 16 bits, since, as presented in figure \ref{fig:distribution} the number of SRAM structures that present more faults is very small, and therefore they do not provide any statistical significance.  Interestingly, both fault injection methods can capture accurately the resiliency of the application to mask errors, since both \textit{correct} lines follow the exact same trend. On the other hand this is not the case for executions in which faults are not masked. Both fault injection methods present the same trend, the more errors that occur the higher the probability for the experiment to result into a crash instead of an SDC. However, in the case of \textit{RND FI} the probability of an error to result into an SDC drops much "faster" in comparison with the probability of the \textit{HW FI}, since the distance between the lines constantly increases as the number of faults increases. The opposite applies for the probability of \textit{Crash}, as the number of errors increase, the \textit{RND FI} probability of crash rapidly increases in comparison to the one of \textit{RND FI}. This effect is not captured in Figure \ref{fig:evalClassification} as the majority of experiments inject two faults, and both fault injection methods behave similarly when injecting a small amount of faults.

In \textit{HW FI} faults present a spatial locality, in other words the distance between any error is much smaller than the distance between errors of the \textit{RND FI}. For example, in \textit{Sobel}, which stores mainly pixel values in a specific cache line, any error corrupting this cache line will result to an \textit{SDC}. In \textit{HW FI} as the number of faults increase the probability to corrupt any cache line in the system does not increase equally, the same cache line or neighboring cache lines (which also store pixel values due to data locality) have a higher probability to have a corruption. Consequently, it is more likely for errors in \textit{HW FI} to corrupt the same application structures and therefore, not deteriorating the outcome of the application even more.  

\subsection{Effect of faults to the Quality of the Output}

Figure \ref{fig:qualities} depicts the quality of the output for each benchmark for the experiments that resulted into the \textit{SDC} category. The quality of the benchmarks in the first row is presented in PSNR and \% Correct Classifications in which the higher the value the better the quality is. The second row of the figure the quality is presented as \textit{Relative Error (RE)} in which the lower the value the better the quality is. Since \textit{RE} can be infinitely large, we bounded it to be up to $1.0$, therefore there are small 'triangles' created on the top of each of the graphs, this is due to the concentration of experiments which present \textit{RE} larger than $1.0$.

Interestingly across all experiments the \textit{HW FI} technique results in outputs with higher quality of the output, except of the \textit{Monte Carlo} benchmark. To be more precise, in \textit{Sobel} ( in Figure \ref{fig:sobelQual} )  the width of the "violin" is broader at higher PSNR values, hence more experiments resulted to outputs with better quality. This is also depicted by the small white dot inside the violins, which represents the average quality of the experiments. \textit{K-Means} demonstrates very high accuracy with both fault injection methods, as the widest part of the 'violin' is observed at the value $100\%$. Consequently, although the algorithm resulted to different cluster centers, during the classification of the observations where assigned to the correct cluster. In any case, the \textit{HW FI} presents once more a wider 'violin' at the highest value, therefore more experiments resulted in better quality values. The same effect can be observed in \textit{DCT, Blackscholes} and in  \textit{Jacobi}. In \textit{Monte Carlo} is the only benchmark which depicts identical qualities among the two fault injection methods, this is due to the randomization of the algorithm, Monte Carlo performs random walks, even if errors occur that corrupt the randomness of the application, the application itself considers these erroneous values as a normal random value. Consequently in the end, the quality of the output for the 2 different fault injection approaches is almost identical. 

\section{Related work}
\label{sec:related}

The reliability implications as well as the energy efficiency of undervolted systems is heavily studied in literature.  Bacha et al.  ~\cite{Bacha:2014:UEF:2742155.2742186,Bacha:2013:DRV:2508148.2485948 } presents an approach which dynamically reduces voltage margins while always preserving safe operation. Their technique is based on the error correction ECC hardware built on modern processors such as the server-class Intel Itanium.  Several approaches propose methods which ensure correct operation of caches under undervolted conditions at the  microarchitectural level~\cite{7551428, Chishti:2009:ICL:1669112.1669126, 4556727}. Architectural techniques  are presented to eliminate data corruption, and by extension enable cache operation at scaled voltage settings. The authors in \cite{8671543,8574581} study the effect of undervotling on several SRAMs of FPGA and they observe the fault patterns and fault location. In our work we study the accuracy of random fault injection models to simulate the behavior of undervolted SRAMs. The authors in \cite{8060425} present a failure model of Near-Threshhold Voltage (NTV) FinFET SRAMs. They propose the use of a Compound Poisson distribution for projecting yield estimates for memory arrays operating in the NTV region. Their model does not take into account the spatial locality that faults present when operating in the NTV region,  In our work we experimentally monitor the effect of the fault spacial locality to the application resiliency and quality of output.  The authors in \cite{Bpredictor} simulate systems with undervolted Branch Predictor Unit and discuss the trade-offs of energy efficiency with the accuracy of the predictor. In this work we focus on undervolting the L1-DCache using real fault maps, and we discuss whether location agnostic fault injection approaches are sufficient to characterize the resiliency of applications. 
\section{Conclusions}
\label{sec:conclusions}
In this paper we present an approach to apply real undervolting SRAM fault maps to a simulated system and observe the resiliency of the applications. We compare the hardware guided fault injection approach with a random guided fault injection approach. There are significant differences in the coarse categorization of the resiliency of the application, which become more obvious as the number of faulty bits increases. There are also differences when inspecting the quality of the output among the two techniques. This is because in an realistic system not all fault locations have the same probability to present faults, therefore from the software perspective the faults can propagate to a limited number of software structures. 

This result does not limit the applicability of random based fault injection. When designing a software fault tolerance technique the designer needs to take into account all possible fault locations, as the technique should be effective regardless of the underlying hardware. However, there is value in optimizing the fault tolerance techniques for a specific system. Similar to compilers, which optimize the source code for the specific hardware, given a set of realistic fault maps that describe the underlying hardware, one can optimize the fault tolerance techniques to be more effective against realistic errors, while being more efficient as the faulty locations are pruned. 

\bibliographystyle{plain}
\bibliography{references}

\end{document}